\documentclass[a4paper,12pt]{article}

\usepackage[T1]{fontenc}
\usepackage{amsmath}
\usepackage{graphicx}

\usepackage{overcite}
\newcommand{\onlinecite}[1]{\hspace{-1 ex} \nocite{#1}\citenum{#1}}

\title{Calculation of Ligand Dissociation Energies in Large Transition-Metal Complexes}
\author{Tamara Husch, Leon Freitag, and Markus Reiher\thanks{corresponding author: markus.reiher@phys.chem.ethz.ch; Phone: +41446334308; Fax: +41446321021.}
\vspace{10 mm}\\
ETH Z\"urich, Laboratorium f\"ur Physikalische Chemie,\\ Vladimir-Prelog-Weg 2, 8093 Z\"urich, Switzerland.
}

\begin{document}

\maketitle

\vspace*{0.3cm}

\begin{center}
{\bf Abstract}
\end{center}

{\small
The accurate calculation of ligand dissociation (or equivalently, ligand binding) energies is crucial
for computational coordination chemistry. Despite its importance,
obtaining accurate {\it ab initio} reference data is difficult and density-functional methods of uncertain
reliability are chosen for feasibility reasons.
Here, we consider advanced coupled-cluster and multi-configurational approaches to
reinvestigate our WCCR10 set of ten gas-phase ligand dissociation energies [J. Chem. Theory Comput. 10 (2014) 3092].
We assess the potential multi-configurational character of all molecules
 involved in these reactions with a multi-reference diagnostic [Mol. Phys. 115 (2017) 2110]
 in order to determine where single-reference
coupled-cluster approaches can be applied. For some reactions of the WCCR10 set,
large deviations from density-functional results including semiclassical dispersion corrections from
experimental reference data had been observed. This puzzling observation deserves special attention here and we tackle the
issue (i) by comparing to {\it ab initio} data that comprise dispersion effects on a rigorous first-principles footing and
(ii) by a comparison of density-functional approaches that model dispersion interactions in various ways.
For two reactions, species exhibiting nonnegligible static electron correlation were identified.
These two reactions represent hard problems for electronic structure methods, also for multi-reference perturbation theories.
However, most of the ligand dissociation reactions in WCCR10 do not exhibit static electron correlation effects, and hence,
we may choose standard single-reference coupled-cluster approaches
to compare with density-functional methods.
For WCCR10, the Minnesota M06-L functional yielded the smallest mean absolute deviation of 13.2\,kJ\,mol$^{-1}$
out of all density functionals considered (PBE, BP86, BLYP, TPSS, M06-L, PBE0, B3LYP, TPSSh, and M06-2X)
without additional dispersion corrections  in comparison to the coupled-cluster results
and the PBE0-D3 functional produced the overall smallest mean absolute deviation of 4.3\,kJ\,mol$^{-1}$.
The agreement of density-functional results with coupled-cluster data
increases significantly upon inclusion of any type of dispersion correction.
It is important to emphasize that
 different density-functional schemes available for this purpose perform equally well.
The coupled-cluster dissociation energies, however, deviate from experimental results on average by 30.3\,kJ\,mol$^{-1}$.
Possible reasons for these deviations are discussed.
}

\vspace*{0.4cm}

\section{Introduction}

A detailed assessment of approximate density functionals
is crucial for drawing meaningful conclusions from quantum chemical studies of transition-metal complexes and their reactions.\cite{Determan2017, Aebersold2017, Ashley2017, Gani2017,Grimmel2016,Swart2016,Kulik2015,Cramer2009,Tekarli2009,j_chem_phys_2006_124_044103,salomon_assertion_2002}
One of the few benchmark sets  for large transition-metal complexes containing experimental gas-phase reference data is our WCCR10 reference set of ligand bonding energies \cite{Weymuth2014}.

The WCCR10 set comprises ten
experimentally measured gas-phase ligand dissociation energies obtained from threshold collision-induced decay (T-CID) experiments.
For nine density functionals,
we found mean absolute errors larger than
$26.7$\,kJ~mol$^{-1}$ with largest absolute errors as large as $83.4$\,kJ~mol$^{-1}$ for the WCCR10 energies. \cite{Weymuth2014}
For most of the ligand dissociation energies,
the subsequent inclusion of semiclassical dispersion corrections yielded \cite{Weymuth2014} results which deviated even further from the experimental
results although they should have brought the results in closer agreement with the experimental data; the reason for this has remained unclear.
In a subsequent study\cite{Weymuth2015}, we attempted to reparametrize the standard
BP86\cite{Perdew1986} functional to achieve better agreement with the experimental
reference energies.
We found \cite{Weymuth2015}, however, that this goal cannot be achieved with a single
parameter set.
So far, results obtained with a new functional reported by the Truhlar group (MN15-L)
agreed best with the experimental WCCR10 data (mean absolute deviation
of 22.8 kJ~mol$^{-1}$). \cite{Yu2016}

Further studies  attempted to elucidate the poor performance of density functional theory (DFT).
Kobylianskii {\it et al.} \cite{Kobylianskii2013} found large
discrepancies between calculated and measured Co--C dissociation energies in two organocobalamins.
Qu {\it et al.} \cite{Qu2015} studied the discrepancies for these cobalamines in more detail
and claimed that the procedure with which the dissociation energies
 are extracted from T-CID experiments cannot be rigorously validated for molecules larger than 50 atoms and the discrepancy might not be
a failure of the theoretical description.
Pollice {\it et al.} \cite{Pollice2017a} attempted to address this issue partially in a later paper, but the discrepancies persisted.

Evidently, the description of the electronic structure of large transition-metal complexes is still a challenging task 
because electronic structures of multi-configurational character may be encountered.
For single-config\-ura\-tional cases, coupled-cluster (CC) methods at the basis-set limit turned out to be very reliable. 
\cite{Minenkov2015a,Sparta2014,Tew2006,Bartlett2010,Klopper2010,Eriksen2016,Tajti2004}
Recent developments such as the domain-based local pair-natural-orbital 
CC (DLPNO-CC) approach of the Neese group\cite{Riplinger2013a,Riplinger2013} or the pair-natural-orbital
local CC (PNO-CC) approach of the Werner
group\cite{Schwilk2017} with explicit correlation factors (PNO-LCCSD-F12)\cite{Ma2017} make CC
calculations feasible for large molecules.
Single-reference CC theory is, however,       
generally not applicable for molecules which exhibit strong static electron correlation.
Unfortunately, decades of research on multi-reference CC theories 
(see, for instance, Refs.\ \onlinecite{Hanauer2011, Evangelista2011,Kohn2013,Jeziorski1981,Hanrath2005,Das2012})
have not led to a unique model that is as well-defined and efficient as its single-reference analogs are.
Consequently, multi-reference perturbation theories, which require a distinction between 
static and dynamic electron correlation, are still a very good option in cases of strong electron correlation. 
Static electron correlation is then captured with a complete active space self-consistent field (CASSCF) ansatz 
\cite{Roos1980,Olsen2011} which requires the selection of a limited number of active orbitals.
This selection, however, can be achieved in a fully automated fashion. \cite{Stein2016, Stein2016a, Stein2017a} 
The density matrix renormalization group (DMRG) \cite{White1992,White1993,Marti2010,Chan2011,Olivares2015,lege08,chan08,chan09,mart11,scho11,kura14,wout14,yana15,szal15,knec16,chan16} ansatz 
provides access to much larger active orbital spaces than those that are
in reach for the standard CASSCF approaches.
The dynamical electron correlation is quantified afterwards by multi-reference perturbation theory,
such as second-order CAS perturbation theory (CASPT2) \cite{Andersson1990,Andersson1992} and 
$N$-electron valence perturbation theory to second-order (NEVPT2). \cite{Angeli2001,Angeli2002,Angeli2001a} 
CASPT2 and NEVPT2 calculations have become feasible for large transition-metal molecules, e.g., through 
Cholesky decomposition (CD) of the two-electron integrals \cite{Roca2011,Aquilante2008CDCASPT2,Aquilante2017,Freitag2017}, 
through the application of density-fitting approaches, \cite{Gyorffy2013}
and through the development of methods applying localized molecular orbitals. \cite{Segarra2015,Guo2016,Segarra2018,Menezes2016}
In general, the efficient implementation of multi-reference perturbation theories that can deal with the 
large active spaces accessible through DMRG has been the subject of intense research in the past decade. 
\cite{Zgid2009,Kurashige2011,Kurashige2014,Sharma2014,Guo2016a,Sharma2016,Roemelt2016,Sokolov2016,Freitag2017}

Here, we investigate the multi-configu\-rational character and the role of static electron correlation in the WCCR10 set
based on DMRG calculations.
In particular, we discuss the suitability of single-reference approaches and then consider
DLPNO-CCSD(T) ligand dissociation energies.
Very recently, Ma {\it et al.} \cite{Ma2017} presented PNO-LCCSD-F12 WCCR10 ligand dissociation
energies, to which we compare our DLPNO-CCSD results.
We then re-assess several
density functionals with a focus on dispersion interactions with reference to DLPNO-CCSD(T).

\section{Computational Methodology}

We adopted BP86\cite{Becke1988, Perdew1986}/def2-QZVPP\cite{Weigend2005} and
 BP86-D3(0)\cite{Grimme2010}/def2-QZVPP optimized structures of all compounds in this work from Ref.\ \onlinecite{Weymuth2014}
to facilitate a direct comparison with our previous work and with recent results published by Ma {\it et al}. \cite{Ma2017,Werner2018}
The molecular structures employed in the multi-configurational calculations were the same as those in the single-configurational calculations.
However, for the calculation of the dissociated complex, unlike in the single-configurational calculations, the two fragment structures were merged
into one structure file with the two fragments separated by 10\,\AA{} to facilitate the selection of the active orbital space.
The ground state for all molecules involved in reactions of the WCCR10 set is
the closed-shell singlet state. 
We did not find any lower-lying open-shell singlet states when calculating the singlet states 
as unrestricted singlets.  
The triplet states are unanimously higher in energy                                                     
than the singlet states (see also Table~16 in the Supporting Information)

We carried out second-order M{\o}ller--Plesset (MP2) perturbation-theory,
spin-component-scaled (SCS-)MP2 \cite{Grimme2003},
and DLPNO-CCSD(T)\cite{Riplinger2013a} coupled-clus\-ter calculations with the \textsc{Orca} program (version 4.0.1) \cite{Neese}.
The frozen-core approximation was invoked for all wave-function-based calculations.
In all MP2 calculations, we exploited resolution-of-the-identity (RI) density fitting.
Auxiliary cc-pVTZ/C or cc-pVQZ/C bases\cite{Weigend2002} were specified
for the RI density fitting in the MP2 calculations.
We denote single-point energies determined with a given method as a first entry in front of a double slash followed by a second
entry for the method employed in the molecular structure
optimization, e.g., DLPNO-CCSD(T)//BP86.
For all MP2, SCS-MP2, and DLPNO-CCSD(T) calculations, we chose a cc-pVTZ or a cc-pVQZ basis set \cite{Dunning1989} for all main group elements
and a cc-pVTZ-PP or a cc-pVQZ-PP basis set \cite{Figgen2005, Peterson2007, Figgen2009} for all transition metals.
Stuttgart effective core potentials\cite{Figgen2005, Peterson2007, Figgen2009} substituted the relativistic core electrons
of all transition-metal atoms.
From the energies obtained with the cc-pVTZ(-PP) basis set and the energies obtained with the cc-pVQZ(-PP) basis set,
we extrapolated to energies at the complete basis set (CBS) limit as proposed in Ref.~\onlinecite{Halkier1998}
(see also the Supporting Information).
We only discuss DLPNO-CCSD(T)/CBS energies in the main text.
All underlying Hartree-Fock calculations were accelerated by the RIJCOSX \cite{Neese2009} approximation
with a corresponding Coulomb fitting basis\cite{Weigend2006}.
All single-point energies were converged to 10$^{-8}$ Hartree
(\textsc{Orca} keyword \texttt{TightSCF}).
We verified the stability of the self-consistently obtained Hartree--Fock solutions 
by drastically perturbing the molecular orbitals as described in Ref.\ \cite{Vaucher2017}.
The results presented in the main text were obtained with \texttt{NormalPNO} thresholds as recommended in
Ref.~\onlinecite{Liakos2015}.

We calculated single-point energies with \textsc{Orca}
with several popular density functionals:
BLYP \cite{Becke1988, Lee1988}, BP86, PBE, \cite{Perdew1996, Perdew1997} M06-L \cite{Zhao2007}, TPSS \cite{Tao2003},
B3LYP\cite{Lee1988,Stephens1994}, PBE0 \cite{Adamo1999}, M06-2X\cite{Zhao2007}, and
TPSSh \cite{Staroverov2003, Staroverov2004},
all with a def2-QZVPP basis set \cite{Weigend2005}
and with a def2-ecp effective core potential \cite{Andrae1990} for Pt, Ru, Au, Ag, and Pd (all other metals were considered in all-electron calculations).
For all pure functionals, the  RI approximation was applied (def2/J fitting basis\cite{Weigend2006}).
In calculations involving hybrid functionals, the RIJCOSX\cite{Neese2009} approximation was activated.

All D3 dispersion (including Axilrod--Teller--Muto contributions \cite{Grimme2010}) were calculated with the stand-alone \textsc{DFT-D3} program\cite{dftd3}
of Grimme and collaborators. We applied the Becke--Johnson damping function for calculating the D3 dispersion contributions for BLYP, BP86, PBE, TPSS, B3LYP, PBE0,
and TPSSh and the zero-damping function for M06-L and M06-2X as recommended in Ref.~\onlinecite{Goerigk2011}.
We abbreviate the semiclassical D3 dispersion corrections with
a zero damping function as D3(0) and the one with a
Becke--Johnson damping function as D3(BJ) \cite{Grimme2011a}.

We considered other types of dispersion corrections in the context of B3LYP calculations:
We chose a nonlocal (NL) van-der-Waals functional which was adapted from
VV10 \cite{Vydrov2010} for the B3LYP exchange-correlation functional \cite{Hujo2011}.
The single-point energies with this functional were evaluated with \textsc{Orca} and the dispersion correction was obtained both self-consistently (B3\-LYP-SCNL) and
non-self-consistently (B3LYP-NL).
We also calculated dispersion corrections with the density-dependent exchange-hole dipole moment (XDM) model
\cite{Becke2005,Becke2006,Angyan2007,Becke2007,Hesselmann2009,
Steinmann2010}.
The required B3LYP/def2-QZVPP electron densities were obtained with the \textsc{Gaussian} program \cite{Gaussian}
and post-processed with the stand-alone \textsc{postg} program \cite{postg}.
The XDM calculations involve two empirical parameters which were set to $a_1=0.6356$ and $a_2=1.5119$,
as recommended for a near-complete basis set \cite{postg}.

We carried out multi-reference CASPT2 calculations.
The IPEA shift \cite{Ghigo2004} was set to zero in the CASPT2 calculations to facilitate comparison with single-reference MP2 results.
To estimate the basis-set effect and deviations from single-reference MP2 calculations, we also calculated MP2 dissociation energies in the ANO-RCC basis set.
Moreover, we carried out \textit{self-consistent-field} DMRG with subsequent strongly contracted NEVPT2 (SC-NEVPT2).
We denote these calculations as DMRG-SC-NEVPT2; note that we omit the "SCF" label for the sake of brevity in this notation.
Accordingly, we denote partially contracted (PC) calculations as DMRG-PC-NEVPT2.

We applied our automated orbital selection protocol\cite{Stein2016, Stein2016a, Stein2017a} for multi-configu\-rational calculations
combined with a multi-configurational diagnostic\cite{Stein2017}, $Z_{s(1)}$, calculated from single-orbital entropies\cite{Legeza2003,Rissler2006}
that were extracted from a qualitatively correct multi-configurational wave function.
The $Z_{s(1)}$ multi-configu\-rational diagnostic\cite{Stein2017} was calculated from the automatedly selected active space, unless noted otherwise.
Due to the very low multi-configurational character of
some compounds and the necessity to have consistent active orbital spaces for
the undissociated complex and its fragments, we adjusted the final orbital selection.
Details on the active-space selection procedure are provided in the Supporting Information.

We chose the all-electron ANO-RCC\cite{Roos2005} basis set with the valence quadruple-zeta polarized (ANO-RCC-VQZP) contraction scheme for the metal atoms and the valence triple-zeta polarized (ANO-RCC-VTZP) contraction scheme for other atoms.
For reaction 4, where multi-reference perturbation theory with this basis set would be computationally unfeasible, and hence, a valence double-zeta polarized (ANO-RCC-VDZP) basis set was chosen.
This mixed basis set is denoted "ANO-RCC" for simplicity in the following.
Two-electron integrals were calculated with the atomic compact Cholesky decomposition (CD) approach\cite{Aquilante2009,Aquilante2008CDMCSCF,Aquilante2008CDCASPT2} with a decomposition threshold of $10^{-4}$.
Scalar-relativistic effects were accounted for through the second-order scalar-relativistic Douglas-Kroll-Hess one-electron Hamiltonian.\cite{hess86,Wolf2002,reih04_2}
All calculations were performed with a locally modified version of \textsc{OpenMolcas} \cite{Aquilante2015}.
For the self-consistent-field DMRG and DMRG-NEVPT2 calculations, we chose our implementations \textsc{QCMaquis} \cite{Dolfi2014,Keller2015,Keller2016,Knecht2016}
and CD-NEVPT2 \cite{Freitag2017},
which are both interfaced to \textsc{OpenMolcas}.
The number of renormalized block states $m$ for all DMRG calculations was chosen such that the truncation error of the DMRG wave function
was less than $10^{-7}$\,a.\,u., which corresponds to $m$=2048 for reaction 9 and $m=$1024 for all other reactions.

For reaction 10, an active space of 17 orbitals was automatedly selected which is, however, too large for a CASPT2 calculation for molecules of this size.
Therefore, we manually selected a smaller active orbital space comprising only nine orbitals for this reaction.
We carried out DMRG-SC-NEVPT2 calculations with the full (17 orbitals) and reduced (9 orbitals) active orbital spaces which resulted in ligand dissociation energies that differed by only 1.8 kJ~mol$^{-1}$.
We consequently chose this reduced active space for reaction 10 for all multi-configurational calculations in this work.

\section{WCCR10 Set of Coordination Energies}

\begin{figure}[ht]
  \centering
  \includegraphics{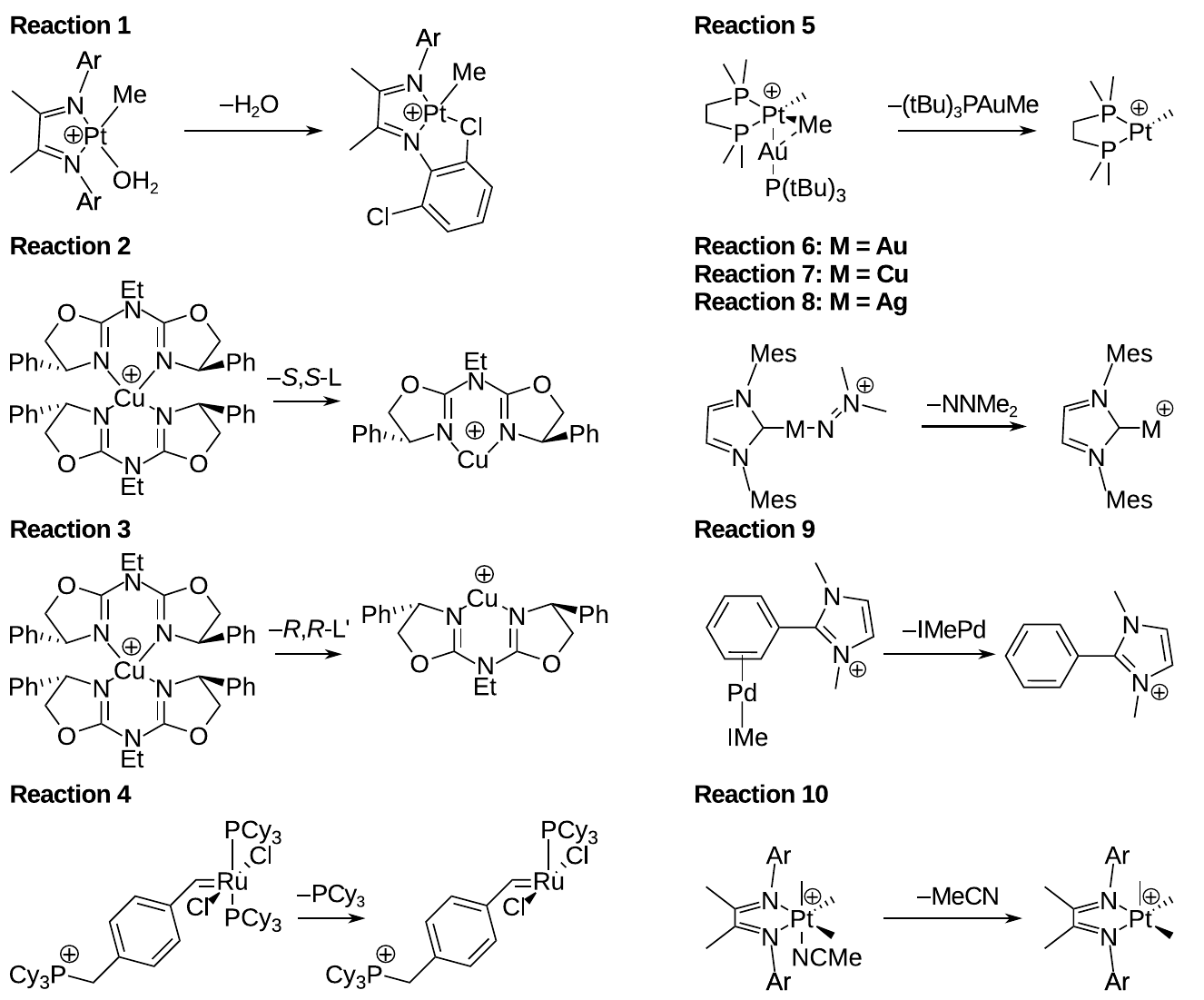}
  \caption{Overview of the ten ligand dissociation reactions in the WCCR10 set.
The abbreviation "Ar" denotes a 2,6-C$_6$H$_3$Cl$_2$ substituent.
The abbreviations "\textit{S},\textit{S}-L" and
"\textit{R},\textit{R}-L'"\, refer to the neutral ligands (L and L')
dissociating from the reactants in reactions 2 and 3, respectively.
}
  \label{fig:wccr10}
\end{figure}

The WCCR10 data set\cite{Weymuth2014} contains ten experimentally determined
ligand dissociation energies of large transition-metal complexes (Figure~\ref{fig:wccr10}).
These cationic complexes feature different transition metals (Au, Ag, Pt, Ru, Cu, Pd) and a
diverse selection of ligand environments.

Experimentally, the ligand dissociation energies were determined from
 T-CID experiments. \cite{Narancic2007, Hammad2005,
Zocher2007b, Torker2008, Weymuth2014, Serra2011, Fedorov2010, Fedorov2010a,
Couzijn2010, Couzijn2014}
The extraction of ligand dissociation energies from these experiments requires
elaborate post-processing. The post-processing of the data includes modeling of ion-molecule collision dynamics,
 approximating a density-of-state function, and a sophisticated fitting procedure. \cite{Narancic2007}
While an uncertainty is attached to each of the experimental
ligand dissociation energies, this uncertainty only accounts for the
uncertainty of one measured quantity and for variations between independent data sets.
It does not account for other uncertainties introduced during data post-processing
such as the application of a single "effective" vibrational frequency when modeling the density-of-states function. \cite{Narancic2007}

We calculate ligand dissociation energies at zero Kelvin as the difference of the
sums of electronic and zero-point vibrational energies (ZPE) for the lowest-energy conformer
of the  reactant and for the two fragments.
DFT ZPEs are known to be a good approximation because they are governed by frequencies of stiff vibrations for which the harmonic approximation works best and because of a fortunate error compensation which makes unscaled BP86 harmonic frequencies calculated with a triple-zeta (or larger) basis set with polarization functions on all atoms in general similar to experimental fundamental
frequencies.\cite{reih2003,Neugebauer2003,reih2004}. Accordingly, we found \cite{Weymuth2014} that
the ZPE correction for reaction 9 determined with different density functionals scatters by
1.5 kJ~mol$^{-1}$. Therefore, for a direct comparison of calculated and experimental dissociation energies, we subtract from  the experimental data the ZPE difference calculated with BP86/def2-QZVPP from Ref.~\onlinecite{Weymuth2014} (see Table~6 in the Supporting Information of
that reference).
Besides the approximation introduced by the electronic structure method itself and the harmonic ZPE contribution,
an important approximation is introduced by relying on a specific molecular structure. In this work, each
compound (reactants and fragments) is only represented by one conformer. This approximation is justified if the dissociation energy is considered at zero
Kelvin, as is the case here,
and each structure actually corresponds to the lowest-energy conformer.
For several of the large structures (in particular, for reactions 2 and 3),
conformational searches were carried out in Ref.~\onlinecite{Weymuth2014} to identify the lowest-energy
conformers.

\section{Results and Discussion}
\subsection{Assessment of Multi-Configurational Character}

Different proposals exist to define a diagnostic of the multi-configurational character 
of a molecule. Recently, we presented the $Z_{s(1)}$ diagnostic \cite{Stein2017} 
as a measure that is obtained from a multi-configurational wave function. 
It is calculated 
from the single-orbital entropies in a partially converged, but qualitatively correct, and 
therefore inexpensive DMRG wave function.
In Ref.~\onlinecite{Stein2017}, we established guidelines regarding the applicability of single-reference
methods based on the $Z_{s(1)}$ value.
In general, single-reference methods will be appropriate when $Z_{s(1)}<0.10$ and 
multi-reference methods will be required when $Z_{s(1)}>0.20$.
Single-reference methods such as CC with sufficiently high excitation 
degree may accurately describe cases which fall in the intermediate regime, $0.10<Z_{s(1)}<0.20$.

A popular measure that is, however, based on a single-reference wave function 
to be evaluated even for a potential multi-configurational case, 
is the $D_1$ diagnostic. \cite{Janssen1998} 
It is calculated from the matrix
norm of the single-excitation amplitude vector of a CC wave
function with single and double excitations.
Janssen \textit{et al.} suggested that $D_1 < 0.02$ indicates single-configurational character and 
$D_1>0.05$ indicates multi-configurational character. \cite{Janssen1998}
In the intermediate regime, $0.02<D_1<0.05$, caution is advised.\cite{Janssen1998} 
We present results for the $Z_{s(1)}$ diagnostic and for the $D_1$ diagnostic 
(obtained from CCSD amplitudes by Werner and collaborators\cite{Ma2017}) in Table~\ref{tab:zs1}.

  \begin{table}
   \centering
   \caption{Multi-configurational $Z_{s(1)}$ diagnostic and 
single-configurational $D_1$ diagnostic from Ref.~\onlinecite{Ma2017} for all
molecules involved in reactions in the WCCR10 set. The reactants and 
products (charged and neutral fragments, respectively) are displayed in Figure~1.}
   \begin{tabular}{cccccc}
     \hline
     \hline
     Reaction & \multicolumn{2}{c}{$Z_{s(1)}$} & \multicolumn{3}{c}{$D_1$ (from Ref.~\citenum{Ma2017})}\\
     & Reactant & Products & Reactant & Charged & Neutral  \\
     &  & &  & Fragment & Fragment \\
     \hline
        1 & 0.15 & 0.15 & 0.028 & 0.028 & 0.017 \\
        2 & 0.09 & 0.07 & 0.028 & 0.029 & 0.028 \\
        3 & 0.09 & 0.10 & 0.028 & 0.029 & 0.028 \\
        4 & 0.22 & 0.37 & 0.044 & 0.032 & 0.021 \\
        5 & 0.12 & 0.04 & 0.023 & 0.022 & 0.027 \\
        6 & 0.17 & 0.12 & 0.030 & 0.030 & 0.041 \\
        7 & 0.12 & 0.15 & 0.031 & 0.037 & 0.041 \\
        8 & 0.14 & 0.14 & 0.031 & 0.037 & 0.041 \\
        9 & 0.23 & 0.12$^{a}$ & 0.034 & 0.027 & 0.035 \\
       10 & 0.14 & 0.15 & 0.028 & 0.029 & 0.030 \\
     \hline
     \hline
\multicolumn{6}{l}{$^{a}$ This diagnostic was evaluated with the final active space employed in} \\
\multicolumn{6}{l}{ the multi-configurational calculation (see also the Supporting Information).} \\
     \hline
   \end{tabular}
\label{tab:zs1}
  \end{table}

The $D_1$ diagnostic indicates that all molecules involved in reactions in the WCCR10 set
fall in the intermediate regime where we may assume that CCSD(T) calculations 
 will yield reliable results ($0.022 < D_1 < 0.044$, see Table~1).
The $Z_{s(1)}$ diagnostic also indicates that the majority of the molecules (fourteen out of twenty species)
fall into an intermediate regime ($0.10<Z_{s(1)}<0.20$, see Table~1).
Three species (reactants of reactions 2 and 3, and products of reaction 2) exhibit a $Z_{s(1)}$ value 
which is slightly lower than the threshold value of $Z_{s(1)}=0.10$, and hence, 
these species can be classified as clear single-configurational cases.
Consequently, multi-reference calculations are generally not 
required to obtain accurate electronic energy differences for eight out of the ten reactions in the WCCR10 set.
The reactants of reactions 4 and 9, and the products of reaction 4 are species 
which exhibit nonnegligible static electron correlation according to the $Z_{s(1)}$
diagnostic ($0.22 < Z_{s(1)} < 0.37$). Hence, reactions 4 and 9 will be interesting 
targets for multi-reference perturbation theories.

\subsection{Ab Initio Dissociation Energies}

\begin{figure}[h]
 \centering
 \includegraphics[width=\textwidth]{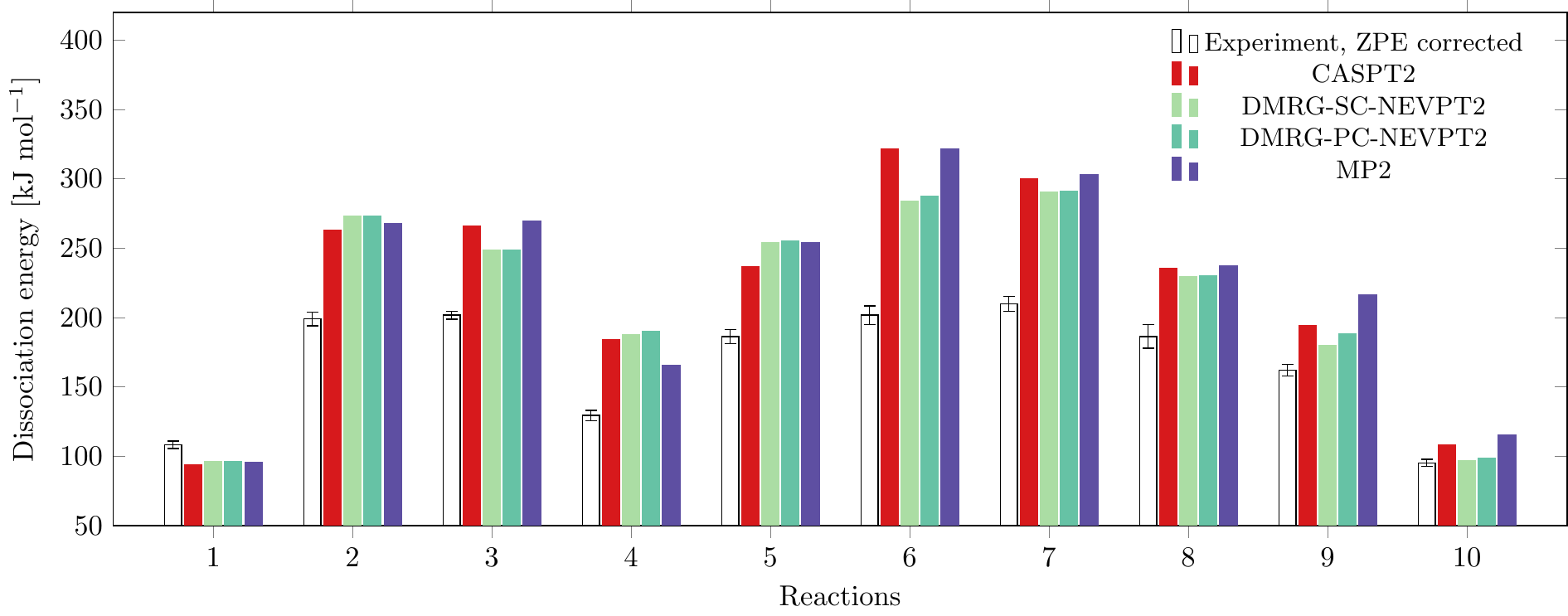}
 \caption{Comparison of ZPE-back corrected experimental and calculated ligand dissociation energies in kJ~mol$^{-1}$.
The ligand dissociation energies were calculated with multi-configurational perturbation theories (CASPT2, DMRG-SC-, and DMRG-PC-NEVPT2) and single-configurational M{\o}ller-Plesset perturbation theory (MP2). All results were obtained with an ANO-RCC basis set.}
\label{fig:mc-energies}
\end{figure}

Figure~\ref{fig:mc-energies} shows the WCCR10 ligand dissociation energies calculated with multi-reference perturbation theories:  CASPT2, DMRG-SC-NEVPT2, and DMRG-PC-NEVPT2
(for numerical data see Table~3 in the Supporting Information).

Given the low amount of static electron correlation for most species in the WCCR10 set,
it is particularly interesting to compare multi-reference perturbation theory with the single-reference methods,
in particular with MP2/ANO-RCC.
For molecules with a low multi-configurational character, we expect the multi-reference
perturbation theories to yield results close to MP2/ANO-RCC,
while a larger deviation is expected for molecules with increased static correlation.
In agreement with this expectation, we observe the largest deviations between the
CASPT2, DMRG-SC-NEVPT2, and DMRG-PC-NEVPT2 and the MP2/ANO-RCC results
for reactions 4 and 9 which are the two reactions for which $Z_{s(1)}>0.20$.
The CASPT2, DMRG-SC-NEVPT2, and DMRG-PC-NEVPT2 results deviate from the MP2/ANO-RCC result by 18.4, 22.0, and 24.4\,kJ~mol$^{-1}$, respectively, for reaction 4 and by 22.2, 36.5, and 28.1\,kJ~mol$^{-1}$ for reaction 9.
For the other reactions, for which $Z_{s(1)}<0.20$,
the deviations between the results obtained CASPT2, DMRG-SC-NEVPT2, and DMRG-PC-NEVPT2
and the MP2/ANO-RCC ligand dissociation energies are on average
 4.9, 13.1, and 12.3\,kJ~mol$^{-1}$, respectively.
The larger deviation of DMRG-SC-NEVPT2 and DMRG-PC-NEVTP2 results from the MP2/ANO-RCC data compared to CASPT2
is due to the Dyall Hamiltonian chosen as the zeroth-order Hamiltonian in NEVPT2. This Hamiltonian includes two-electron interactions in the active space. Therefore, it is capable of recovering more correlation energy already at zeroth order.
The application of an IPEA shift of 0.25 a.u.\ for CASPT2 calculations leads, on average, to an 
increase of the ligand dissociation energies by about 4.5 kJ~mol$^{-1}$ (see Table~3 in the Supporting Information).

\begin{table}[ht]
 \caption{ZPE-back corrected
experimental ligand dissociation energies and DLPNO-CCSD(T) ligand dissociation energies in kJ~mol$^{-1}$.
All DLPNO-CCSD(T) calculations were extrapolated to the complete basis set limit
from single-point energies obtained with a cc-pVTZ(-PP) and a cc-pVQZ(-PP) basis set,
the single-point calculations were carried out for BP86/def2-QZVPP and BP86-D3(0)/def2-QZVPP
optimized structures. }
 \centering
 \begin{tabular}{lrlrr}
  \hline
  \hline
  Rct. & \multicolumn{1}{c}{Experiment}  & Reference
& \multicolumn{1}{c}{DLPNO-CCSD(T)/}
& \multicolumn{1}{c}{DLPNO-CCSD(T)/}  \\
   &   &
& \multicolumn{1}{c}{CBS//BP86/}
& \multicolumn{1}{c}{CBS//BP86-D3(0)/}  \\
   &   &
& \multicolumn{1}{c}{def2-QZVPP}
& \multicolumn{1}{c}{def2-QZVPP}  \\
  \hline
1  & 108.3  $\pm$ 2.7 &\onlinecite{Hammad2005}                & 105.6 & 107.1 \\
2  & 199.1  $\pm$ 5.0 &\onlinecite{Zocher2007b}               & 238.2 & 265.1 \\
3  & 201.8  $\pm$ 2.9 &\onlinecite{Zocher2007b}               & 237.7 & 263.6 \\
4  & 129.4  $\pm$ 3.8 &\onlinecite{Torker2008, Weymuth2014}   & 207.0 & 220.5 \\
5  & 186.3  $\pm$ 5.0 &\onlinecite{Serra2011, Weymuth2014}    & 183.7 & 188.6 \\
6  & 201.8  $\pm$ 6.7 &\onlinecite{Fedorov2010, Fedorov2010a} & 278.4 & 281.9 \\
7  & 209.8  $\pm$ 5.4 &\onlinecite{Fedorov2010, Fedorov2010a} & 244.5 & 250.0 \\
8  & 186.4  $\pm$ 8.4 &\onlinecite{Fedorov2010, Fedorov2010a} & 205.9 & 209.7 \\
9  & 162.0  $\pm$ 4.2 &\onlinecite{Couzijn2010}               & 151.0 & 152.2 \\
10 & 95.3   $\pm$ 2.5 &\onlinecite{Couzijn2014}               &  98.5 &  99.5 \\
  \hline
  \hline
 \end{tabular}
 \label{tab:ccsdResults}
\end{table}
Table~\ref{tab:ccsdResults} additionally presents WCCR10 ligand dissociation energies
obtained with DLPNO-CCSD(T) and the experimental results.
On average, the DLPNO-CCSD(T)//BP86 results differ by 30.3 kJ~mol$^{-1}$ from the experimental data and individual deviations may be as large as 77.6 kJ~mol$^{-1}$ (reaction 4).
The optimization of the WCCR10 structures with several density functionals without any dispersion correction resulted in nearly
identical structures. \cite{Weymuth2014} We, therefore, do not expect large effects on
CC energies for those structures.
We, however, found sizable differences between structures optimized with ordinary density functionals and structures optimized with dispersion-cor\-rected density functionals. \cite{Weymuth2014}
Hence, we additionally evaluated DLPNO-CCSD(T) ligand dissociation energies for BP86-D3(0) optimized structures (see last column
in Table~\ref{tab:ccsdResults}).
The disagreement with the experimental data increased compared to DLPNO-CCSD(T)//BP86: the mean absolute deviation of DLPNO-CCSD(T)//BP86-D3(0) complexation energies with respect to the
experimental data is 38.0 kJ~mol$^{-1}$ and the largest absolute deviation is 91.1 kJ~mol$^{-1}$ (reaction 4).
For all reactions, the DLPNO-CCSD(T)//BP86-D3(0) ligand dissociation energies are larger than the DLPNO-CCSD(T)//BP86
results. We observe the largest difference between the DLPNO-CCSD(T)//BP86-D3(0) and DLPNO-CCSD(T)//BP86 ligand dissociation energies for reactions 2, 3, and 4 (26.9, 25.9, and 13.5 kJ~mol$^{-1}$, respectively).
We should emphasize that one would expect that dispersion-corrected molecular structures should match well with those
present in the gas phase of the mass spectrometer in which the experimental results were obtained. \cite{Grimme2013,Steinmetz2014}

In summary, the ligand dissociation energies calculated with DLPNO-CCSD(T)
agree well with the experimental data for reaction 1, 5, and 10, while they are at least 19.4 kJ~mol$^{-1}$ larger than the experimental ones for reactions 2, 3, 6, 7, and 8.
Although reactions 1 and 10 feature small complexes and fragments, a comparison to reactions 6--8 shows that size cannot be considered a decisive cause for the good agreement.
Our findings parallel those of Qu {\it et al.} \cite{Qu2015} and Pollice {\it et al.} \cite{Pollice2017a} who reported nonnegligible deviations of DLPNO-CCSD(T) ligand dissociation energies from
results for different molecules.

We now turn our attention to the reaction with the largest $Z_{s(1)}$ measures, reactions 4 and 9.
For reaction 9, the DLPNO-CCSD(T) ligand dissociation energy is by 11.0 kJ~mol$^{-1}$ \textit{smaller} than the
measured energy. The CASPT2, DMRG-SC-NEVPT2, and DMRG-PC-NEVPT2 results are, by contrast, 32.0, 17.7, and
26.1 kJ~mol$^{-1}$ \textit{larger} than the experimental value, respectively.
For reaction 4, we observe a very large difference of 77.6 kJ~mol$^{-1}$ between the measured and the DLPNO-CCSD(T) energy.
The CASPT2, DMRG-SC-NEVPT2, and DMRG-PC-NEVPT2 dissociation energies differ from the experimental data by
54.4, 58.0, and 60.4 kJ\,mol$^{-1}$, respectively.
Unfortunately, reaction 4 is the reaction for which we were restricted to
a valence double-zeta polarized basis set for non-metal atoms due to the large size of the reactant.
From the comparison of MP2/ANO-RCC and MP2/CBS results, we can estimate the remaining basis set effect to be
 40.9 kJ~mol$^{-1}$ which is very large.
We may conclude (although only cautiously for reaction 4) that for reactions 4 and 9, neither DLPNO-CCSD(T) nor the multi-reference perturbation theories yield satisfactory results.
Reactions 9 and 4 might, hence, be valuable targets for multi-reference CC approaches at the basis-set limit.

\subsection{Assessment of Density Functionals and Single-Ref\-er\-ence Perturbation Theories}
\label{subsec:benchmarking}

We now consider the accuracy of more approximate single-reference electronic structure models by comparing to the DLPNO-CCSD(T) ligand dissociation energies as a reference.
All calculations were carried out for the BP86/def2-QZVPP optimized structures.
Table~\ref{tab:errorStats} collects the mean absolute and largest absolute deviations (MAD and LAD, respectively)
of ligand dissociation energies calculated with  a selection of density functionals
(PBE, BP86, BLYP, TPSS, M06-L, PBE0, B3LYP, TPSSh, and M06-2X)
with and without D3 dispersion corrections, MP2, and SCS-MP2 with respect to the DLPNO-CCSD(T) results (see the Supporting Information for the numerical data).

\begin{table}[h!]
 \caption{Mean absolute deviations (MAD) and largest absolute deviations (LAD)
of ligand dissociation energies calculated with various approximate electronic
structure models with respect to DLPNO-CCSD(T) data in kJ~mol$^{-1}$.
We indicate in parenthesis for which reaction the LAD was found.
All DFT calculations were carried out with a def2-QZVPP basis set.
We extrapolated cc-pVTZ(-PP) and cc-pVQZ(-PP) MP2 and SCS-MP2 results
 to the complete basis set limit.}
 \centering
 \begin{tabular}{lrrc}
  \hline
  \hline
  Method & MAD & LAD & \\
  \hline
PBE         & 32.5 &  96.3 & (Rct. 4)\\
BP86        & 41.2 & 113.2 & (Rct. 4)\\
BLYP        & 56.1 & 146.4 & (Rct. 4)\\
TPSS        & 34.4 &  97.3 & (Rct. 4)\\
M06-L       & 13.2 &  28.6 & (Rct. 6)\\
PBE0        & 29.7 &  74.3 & (Rct. 4)\\
B3LYP       & 47.0 & 112.2 & (Rct. 4)\\
TPSSh       & 32.1 &  80.7 & (Rct. 4)\\
M06-2X      & 21.4 &  41.3 & (Rct. 6)\\
PBE-D3(BJ)  & 10.2 &  23.6 & (Rct. 7)\\
BP86-D3(BJ) & 17.3 &  31.5 & (Rct. 5)\\
BLYP-D3(BJ) &  9.2 &  17.5 & (Rct. 6)\\
TPSS-D3(BJ) & 10.6 &  25.0 & (Rct. 7)\\
M06-L-D3(0) &  7.7 &  25.4 & (Rct. 6)\\
PBE0-D3(BJ) &  4.3 &   9.1 & (Rct. 9)\\
B3LYP-D3(BJ)&  8.2 &  15.8 & (Rct. 3)\\
TPSSh-D3(BJ)& 10.8 &  21.1 & (Rct. 5)\\
M06-2X-D3(0)& 16.3 &  38.3 & (Rct. 6)\\
SCS-MP2/CBS     & 16.8 &  29.7 & (Rct. 4)\\
MP2/CBS         & 29.3 &  66.2 & (Rct. 9)\\
  \hline
  \hline
 \end{tabular}
 \label{tab:errorStats}
\end{table}

\paragraph{Density functionals without dispersion correction.}

We first turn our attention to the results obtained with the PBE, BP86, BLYP, TPSS, PBE0, B3LYP, and TPSSh functionals.
We observe the largest absolute deviation of the ligand dissociation energies calculated with any of these functionals
 to the DLPNO-CCSD(T) results for reaction 4.
The ligand dissociation energy of reaction 4 is
underestimated by 74.3 kJ~mol$^{-1}$ (PBE0) to 146.4 kJ~mol$^{-1}$ (BLYP, LADs in Table~\ref{tab:errorStats}).
In fact, the ligand dissociation energies for at least nine out of the ten reactions are smaller than the DLPNO-CCSD(T)
results
when they are calculated with one of these functionals, and overall, the obtained results deviate strongly from the DLPNO-CCSD(T) data
(MAD $> 29.7 $ kJ~mol$^{-1}$, LAD $> 74.3 $ kJ~mol$^{-1}$).
An unsatisfactory agreement of the ligand dissociation energies determined with
non-dispersion-corrected density functionals with DLPNO-CCSD(T) energies
is, however, expected due to the lack of attractive dispersion interactions in the (undissociated) complexes.

The Minnesota functionals M06-L and M06-2X
 describe dispersion interactions to some degree by parametrization of a flexible functional form. \cite{Zhao2007}
Accordingly, the deviation of the ligand dissociation energy for reaction 4 from the DLPNO-CCSD(T) result only amounts to 26.1 kJ~mol$^{-1}$ for M06-L and to 18.1 kJ~mol$^{-1}$ for M06-2X (compared to at least 74.3 kJ~mol$^{-1}$ for  one of the other functionals).
For the complete WCCR10 set, the meta-GGA functional M06-L exhibits the highest accuracy amongst all density functionals without dispersion corrections with an MAD of 13.2 kJ~mol$^{-1}$, while M06-2X achieves an MAD of 21.4 kJ~mol$^{-1}$.
M06-L and M06-2X, however, still lead to too small ligand dissociation energies for nine out of the ten reactions in comparison to DLPNO-CCSD(T) (with the exceptions of reaction 7 (M06-L) and reaction 1 (M06-2X)).
M06 functionals do not reproduce the proper scaling of dispersive interactions with
the sixth inverse power of the distance \cite{Goerigk2011}, and hence, still underestimate the dispersion interactions for most reactions.

\paragraph{ M{\o}ller-Plesset perturbation theories.}

In general, ordinary MP2/CBS does not perform significantly better than most density functionals for the WCCR10 set (MAD = 29.3 kJ~mol$^{-1}$, LAD = 66.2 kJ~mol$^{-1}$).
In fact, MP2/CBS overestimates the ligand dissociation energies for all reactions but reaction 1 which is underestimated by 15.4 kJ~mol$^{-1}$.
For reactions 2--10, the MP2/CBS ligand dissociation energies are on average 30.8 kJ~mol$^{-1}$ larger than the DLPNO-CCSD(T)
results.
SCS-MP2 \cite{Grimme2003} corrects partially for shortcomings of MP2 by scaling the same-spin
and opposite-spin components in MP2 differently.
For the complete WCCR10 set, SCS-MP2/CBS achieves an MAD of 16.8 kJ~mol$^{-1}$ and
an LAD of 29.7 kJ~mol$^{-1}$.
Hence, the overall reliability of SCS-MP2/CBS is worse than what we found for the M06-L functional,
but is still on par with several dispersion-corrected density functionals.

\paragraph{Dispersion-corrected density functionals.}

The application of dispersion corrections in DFT has become a \textit{de facto} standard
and we investigate several dispersion corrections in more detail.
A popular way to account for dispersion interactions in DFT calculations is the application of
semiclassical D3 dispersion corrections\cite{Grimme2010}.
The inclusion of D3 corrections leads to an increase of the predicted ligand dissociation energies
in comparison to the ones calculated without D3 corrections for all reactions in the WCCR10 set and for all functionals.
The resulting B3LYP-D3(BJ) ligand dissociation energies agree better with the DLPNO-CCSD(T) energies
than the B3LYP ones for every single reaction.
For the PBE0, BLYP, PBE, BP86, TPSS, and TPSSh functionals, the agreement is improved for at least seven out of the ten reactions when invoking D3 corrections.
These improvements may be enormous in some cases, e.g., the deviation from the DLPNO-CCSD(T) result for reaction 4 decreases between 59.7 kJ~mol$^{-1}$ (TPSSh vs. TPSSh-D3(BJ)) and 145.9 kJ~mol$^{-1}$ (BLYP vs. BLYP-D3(BJ)) when including D3 corrections compared to the one for the uncorrected functionals.
For M06-L and M06-2X, we also observe a decrease of the deviations for at least nine out of the ten reactions when including D3 corrections.
Note, however, how the M06 series of functionals show similar MAD with and without the D3 dispersion correction
owing to the fact that the uncorrected functionals already include dispersive interactions to some degree
through their parametrization.
For the complete WCCR10 set, the best agreement with the DLPNO-CCSD(T) data was found for PBE0-D3(BJ) (MAD = 4.3 kJ~mol$^{-1}$, LAD = 9.1 kJ~mol$^{-1}$). Two other dispersion-corrected functionals (BLYP-D3(BJ), B3LYP-D3(BJ)) yield MAD's below 10 kJ~mol$^{-1}$ and LAD's below 20 kJ~mol$^{-1}$.
In general, the mean and largest absolute deviations of the results obtained with dispersion-corrected density functionals are smaller than the ones obtained for their non-dispersion-corrected counterparts for all functionals considered and do not exceed 17.3 kJ~mol$^{-1}$ and 38.3 kJ~mol$^{-1}$, respectively (MAD of BP86-D3(BJ) and LAD of M06-2X-D3(0), see also Table~\ref{tab:errorStats}).

The D3 corrections incorporate fit parameters which introduce a prediction uncertainty. \cite{Weymuth2018}
We estimate the uncertainties of the B3LYP-D3(BJ) dispersion energies with the \textsc{BootD3} program. \cite{Weymuth2018}
As reported in Ref.\ \onlinecite{Weymuth2018}, the uncertainty of 
(absolute) dispersion energies grows with molecular 
size. Not surprisingly, we observe the largest uncertainty of an absolute dispersion energy for the largest 
molecule in the WCCR10 set, namely for the reactant of reaction 4 with an absolute B3LYP-D3(BJ) dispersion 
energy of $-1260.4$ kJ~mol$^{-1}$ with a standard deviation of 49.6 kJ~mol$^{-1}$ (see Table~18 in the 
Supporting Information).
The uncertainty of the dispersion contributions to the B3LYP-D3(BJ) reaction energies is, however, much smaller, i.e.,
less than 2.7 kJ~mol$^{-1}$ for eight out of the ten reactions.
Larger uncertainties we found for reactions 4 and 5
(4.5 and 6.6 kJ~mol$^{-1}$, respectively).

\begin{figure}[ht]
  \centering
  \includegraphics{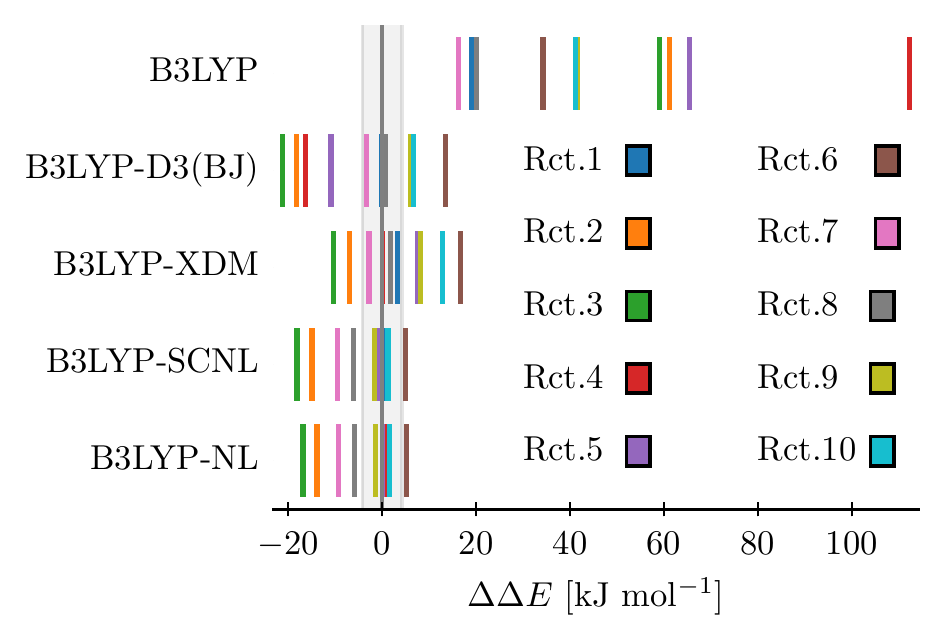}
  \caption{Deviation of electronic ligand dissociation energies $\Delta E$ in kJ~mol$^{-1}$ calculated with (dispersion corrected) B3LYP from DLPNO-CCSD(T) values.
All (dispersion-corrected) B3LYP
energies were calculated with a
def2-QZVPP basis set for BP86/def2-QZVPP optimized structures.
The gray region indicates the region where chemical accuracy is achieved, i.e., where deviations from the
DLPNO-CCSD(T) data are smaller than 4.2 kJ~mol$^{-1}$.}
  \label{fig:threadPlotDB3LYP}
\end{figure}

Furthermore, the D3 dispersion corrections only depend on atom positions, but not on the electron density 
(or even the charge) of the molecule (see, e.g., Ref.~\onlinecite{Hansen2014}).
Transition-metal complexes can, however, show a variety of low-lying electronic states for similar molecular
structures (e.g., in case of different spin states).
We therefore also investigated other types of dispersion corrections for the B3LYP functional: XDM, NL, and SCNL (see Figure~\ref{fig:threadPlotDB3LYP}).

The results obtained with B3LYP-SCNL and B3LYP-NL differ at most by 1.7 kJ~mol$^{-1}$ (reaction 4)
for the reactions in the WCCR10 set.
Apparently, the electron density is not affected severely by the inclusion of nonlocal dispersion corrections
of this type.
Figure~\ref{fig:threadPlotDB3LYP} illustrates that the inclusion of any type of dispersion correction to the B3LYP ligand dissociation energies leads to an improved agreement with the DLPNO-CCSD(T) results for every single reaction.
Nevertheless, we still encounter sizeable deviations. The deviations of B3LYP-D3(BJ) ligand dissociation energies from DLPNO-CCSD(T) energies exceed 11.6 kJ~mol$^{-1}$ for reactions 2, 3, 4, and 6.
Similarly large errors are encountered for reactions 2 and 3 when the ligand dissociation energies are calculated with
B3LYP-SCNL (14.9 and 18.1 kJ~mol$^{-1}$, respectively).
The B3LYP-XDM ligand dissociation energies tend to be too small so that the ligand dissociation energies
for reactions 6 and 10 are underestimated by 16.7 and 12.8 kJ~mol$^{-1}$, respectively.
Overall, the MADs of B3LYP-D3(BJ), B3LYP-NL, B3LYP-SCNL, and B3LYP-XDM results with respect to the DLPNO-CCSD(T) ligand dissociation energies are
between 5.6 kJ~mol$^{-1}$ (B3LYP-NL) and 8.2 kJ~mol$^{-1}$ (B3LYP-D3(BJ)) compared to an MAD of 47.0 kJ~mol$^{-1}$ for the dispersion-uncorrected B3LYP results.

\section{Possible Origins of Discrepancies between Calculated and Experimental Ligand Dissociation Energies}

At present, it is difficult to shed more light on the source of discrepancies between theory and experiment as neither
is without limitations. However, in view of the different quantum chemical approaches considered in this
paper, we may draw some conclusions after having put together more pieces of the puzzle.

First of all, we emphasize that the two other approaches (XDM and NL) to incorporate
dispersion corrections into dispersion-free density functionals yield similar
results compared to
the semiclassical dispersion corrections by Grimme
and to CC calculations
which was also noted for other transition-metal complexes. \cite{Paenurk2017}

The correlation energy in CC calculations is known to converge only slowly
with the basis-set size which can introduce sizable errors.
Ma {\it et al.} \cite{Ma2017} obtained PNO-LCCSD-F12b/VTZ-F12 ligand dissociation energies
for the WCCR10 set which are well converged with respect to the
basis-set size; the deviation of PNO-LCCSD-F12b/VTZ-F12
from PNO-LCCSD-F12b/VDZ-F12 results does not exceed 2.9\,kJ~mol$^{-1}$.
Our DLPNO-CCSD results deviate on average by 4.8 kJ~mol$^{-1}$ 
(at most by 8.2 kJ~mol$^{-1}$ for reaction 2, Table~5 in the Supporting Information) from the
PNO-LCCSD-F12b/VTZ-F12 results which indicates that the former
are converged reasonably well with respect to the basis-set size.
A larger basis set would, however, be desirable especially for reactions 2, 3, and 4.

The comparison with PNO-LCCSD(T)-F12b/VTZ-F12 data \cite{Werner2018} highlights an issue associated with 
 DLPNO-CCSD(T) calculations.
While our DLPNO-CCSD(T) data deviate by less than 1.6 kJ~mol$^{-1}$ from PNO-LCCSD(T)-F12b energies 
for six out of the ten reactions, larger deviations of up to 18.0 kJ~mol$^{-1}$ were found for reaction 7.
We can attribute these differences to the applied PNO thresholds; the deviations 
for reactions 1, 7, and 9 decrease from 5.9, 18.0, and 5.0 kJ~mol$^{-1}$
to 1.8, 10.9, and 0.1 kJ~mol$^{-1}$, respectively, when \textsc{TightPNO} settings are applied in the 
DLPNO-CCSD(T) calculations (see Table~7 in the Supporting Information).
Evidently, even tighter PNO thresholds would be desirable for reaction 7.
The quadruple-zeta basis set and \textsc{TightPNO} settings are already very costly in terms of computing time.
This highlights the benefits of explicitly correlated methods which 
 converge faster with respect to the basis-set size.
We emphasize that the deviations to the experimental data also persist when comparing against 
PNO-LCCSD(T)-F12b results \cite{Werner2018}
and that both methods, DLPNO-CCSD(T) and PNO-LCCSD(T)-F12, were found to
generate much smaller deviations in other reference data sets.\cite{Minenkov2015a,Riplinger2013a,Riplinger2013,Sparta2014,Schwilk2017, Ma2017,Liakos2015}

Apart from intrinsic limitations of electronic structure methods,
we must keep in mind that the molecular structure model is a possible cause for the origin of discrepancies between theory and experiment.
Different neutral dissociation products are, e.g., possible for reactions 6--8 (e.g., 1,1-dimethyldiazene (as assumed in this study), cis-1,2-dimethyldiazene, trans-1,2-dimethyldiazene, or formaldehyde methylhydrazone)
that convert into each other through unimolecular re-arrangements of methyl groups. \cite{Fedorov2010, Fedorov2010a}
It cannot (and does not need to) be determined by mass spectrometry which of these products is formed.
A choice for one structures is, however, required to carry out electronic structure calculations.
The most likely dissociation product for reactions 6--8 was identified in DFT calculations \cite{Fedorov2010, Fedorov2010a}
 which may present issues according to our work in Ref.\ \onlinecite{Weymuth2014} and Section~\ref{subsec:benchmarking}.
Additionally, the consideration of structures which do not correspond to the lowest-energy conformers may be a source of error.
Although, for the structures reported in Ref.\ \onlinecite{Weymuth2014}
conformational sampling was considered to a certain degree (see the Supporting Information of that paper), further investigations might be appropriate.

Furthermore, modeling assumptions exist in the
post-processing protocol of the experimental data.
Several factors in this post-processing protocol were
discussed \cite{Narancic2007, Torker2008, Fedorov2010, Fedorov2010a, Couzijn2014, Zocher2007b}
which possibly affect the experimentally determined ligand dissociation energies
for the reactions in the WCCR10 set as we shall review in the following.

The experimental ligand dissociation energies vary slightly with the number of assumed internal
rotors (input parameter).\cite{Narancic2007}
Although this protocol was only verified for ions with up to fifty atoms, \cite{Narancic2007}
it is the better fulfilled, the larger the dissociating molecules become.
This is due to the fact that the ratio of the moments of inertia between reactant and dissociation products and consequently the rigid-rotor partition functions tends to one for large reactants.
For reaction 4, the largest and smallest ligand dissociation energies determined with different numbers of internal rotors
vary\cite{Torker2008} by 10 kJ~mol$^{-1}$ which is much smaller than the deviation of 77.6 kJ~mol$^{-1}$
between the experimental and the DLPNO-CCSD(T) data.

The post-processing strategy necessitates a choice between a "tight" and a "loose" transition-state
model. \cite{Narancic2007}
A tight transition-state model is assumed to represent reactions well in which an intramolecular
rearrangement is rate limiting, whereas a loose transition-state model is appropriate
when a dissociation process determines the rate.
The overall dissociation reaction may, however, involve multiple steps and
it is not \textit{a priori} clear whether the rate-limiting step
is the ligand dissociation reaction itself or, e.g.,
an intramolecular rearrangement preceding ligand dissociation.
Hence, prior knowledge is required to choose either the loose or the
tight transition-state model. \cite{Couzijn2014}
Generally, the experimental ligand dissociation energy turns out to be larger when assuming a loose transition state than when assuming a tight one.
For all reactions in the WCCR10 set, the loose transition-state model was chosen during the post processing yielding a maximum value for the dissociation energy.
Therefore, the large observed deviations between experimental and CC data cannot be explained
by the choice of transition-state model because the CC results are even larger than the experimental energies for the reactions for which we observe large discrepancies (reactions 2, 3, 6, 7, and 8).
Hence, the differences between measured and calculated energies would become even larger when the tight transition-state model would be selected in the post-processing step.

In addition, further
sources of uncertainties in the post-processing protocol may exist.
For instance, the reactant is thermalized at 343 K \cite{Narancic2007} which is going to result in the formation of
a conformational equilibrium. This implies that the reactant may collide with the collision gas
in various conformations and the reported ligand dissociation energies represent
some sort of average over the accessible conformations
(even if fast energy redistribution and relaxation may populate only few lowest-energy conformers).

To address the reliability of the post-processing protocol for large complexes, 
Pollice \textit{et al.}\cite{Pollice2017a} devised an example which
is largely independent of the assumptions in this protocol.
These authors synthesized a proton-bound dimer which can undergo two alternative dissociation reactions
(hydrogen bond cleavage or O-NO bond cleavage) upon collision (see Figure~4).
Only the products which result from a hydrogen bond
cleavage reaction were observed in the mass spectrometer.
The DLPNO-CCSD(T) bond dissociation energy for the 
hydrogen bond cleavage was, however, found to be larger than the one for the O-NO bond cleavage from which they concluded that 
 the wrong product is predicted.\cite{Pollice2017a}
This experimental finding puts an unexpectedly large deviation of at least 21 kJ~mol$^{-1}$
on the calculated DLPNO-CCSD(T) dissociation energies.
However, we should also note that the theoretical prediction of a major product requires the 
 determination of free energy differences at the respective temperature
which has recently been emphasized by Carpenter and co-workers. \cite{Carpenter2017}
Carpenter \textit{et al.} studied the dissociation of a para-nitrobenzylpyridinium cation in T-CID experiments. \cite{Carpenter2017}
The bond dissociation energy associated with the reaction yielding the major product was larger than the one 
associated with the reaction yielding the minor product. 
Carpenter \textit{et al.} rationalized why the major product was observed  
by including the entropies associated with the dissociation processes in their analysis. \cite{Carpenter2017}
\begin{figure}[ht]
  \centering
  \includegraphics[scale=1.2]{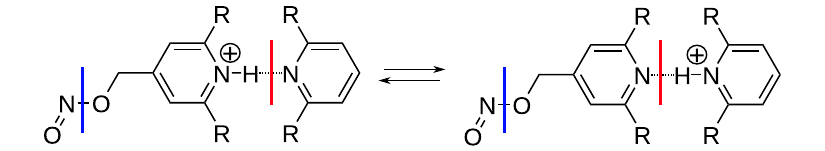}
  \caption{Lewis structures of a proton-bound dimer (R denotes 3,5-di-\textit{tert}-butyl-phenyl substituents) which can undergo either homolytic O-NO bond cleavage
(blue line) or hydrogen-bond cleavage (red line).}
  \label{fig:illustrationPollice}
\end{figure}

Furthermore, the comparison of the theoretical dissociation pathways with the experimental result assumes full statistical
intramolecular vibrational energy redistribution (IVR). This assumption is fulfilled only if the dissociation occurs on a slower timescale than IVR. If the dissociation competes with IVR, a ligand with a larger collision cross-section
is more likely to dissociate.
Dissociation reactions do not compete with IVR in general, but cases in which a competition occurs have been reported for small molecules \cite{Turecek1984, Turecek2003} and medium-sized non-covalently bound ions. \cite{Shaffer2012}
For the example in Figure~\ref{fig:illustrationPollice}, the collision cross-section of the
pyridine monomer dissociating as a result of the hydrogen bond cleavage might be much larger
than that of NO, and hence, hydrogen bond cleavage could be preferred.
However, the experimental set-up in Refs.~\onlinecite{Turecek1984} and \onlinecite{Turecek2003}
promotes fast reactions and non-ergodic behavior,
which is not necessarily the case for the comparatively long-lived collision complexes produced in T-CID experiments.

To conclude, although we have carefully re-examined possible sources of uncertainty in the electronic structure models (see also the Supporting Information) and in experiment,
none of the usual suspects appear to be accountable for the large deviation of high-level CC data from the experimental energies.

\section{Conclusions} 

We revisited the WCCR10 set of ten ligand dissociation energies to provide
results from correlated single- and multi-reference wave function methods and to further investigate the role of dispersion interactions.

We first assessed the multi-configurational character of all molecules in the WCCR10 set with our
multi-reference $Z_{s(1)}$ diagnostic which is based on a qualitatively correct multi-configurational wave function.
Two reactions (reactions 4 and 9) turned out to involve molecules which exhibit nonnegligible static electron correlation.
Our results showed that these two reactions were even challenging for multi-reference perturbation theories
and satisfactory agreement with the experimental data was not achieved for either reaction.
These two reactions therefore represent interesting targets for other multi-reference methods such as
multi-reference coupled-cluster approaches at the basis-set limit. 

After ascertaining that single-reference approaches are adequate for eight out of the ten reactions in the 
WCCR10 set, we carried out DLPNO-CCSD(T) calculations. We were
able to achieve a reasonable agreement with the experimental data for three reactions (reactions 1, 5, and 10).
For the other five reactions (reactions 2, 3, 6, 7, and 8),
which involve single-configurational complexes and fragments according to the Z$_s(1)$
diagnostic, we observed deviations between 19.4 and 76.6 kJ~mol$^{-1}$ from the experimental energies.
We ensured that our coupled-cluster results were essentially converged with respect to the basis-set size
by comparison of our DLPNO-CC data to explicitly correlated CC results from Refs.\ \onlinecite{Ma2017, Werner2018}.
The comparison of our DLPNO-CCSD(T) data with explicitly correlated CCSD(T) data from Ref.\ \onlinecite{Werner2018}
showed that the application of tight PNO thresholds might be required for accurate results in some 
cases (reaction 7).
We also addressed the structural uncertainty by comparing results obtained for BP86 and BP86-D3(0) structures.
Contrary to ones' expectation, the deviation of the DLPNO-CCSD(T) results from the experimental data increased even
further when BP86-D3(0) instead of BP86 structures were chosen.
Neither the basis-set size nor the threshold criteria nor the structural uncertainties appear to be able to account
for the very large discrepancies between coupled-cluster and experimental data.
While we discussed several possible causes for the observed deviations, it is currently not
possible to pinpoint the cause because there is simply not enough data available to detect trends.
Hence, it is indispensable to consider more experimental gas-phase data on binding energies of
medium-sized to large molecules in order to
finally track down the sources of these nagging discrepancies.

For the single-configurational molecules, we compared the results of density-functional calculations and single-reference
perturbation theories with DLPNO-CCSD(T) data. 
We generally cannot recommend the application of single-reference perturbation theories in this context
because the overall reliability of MP2 and SCS-MP2
(MAD of 29.3 kJ~mol$^{-1}$ and 16.8 kJ~mol$^{-1}$, respectively)
was worse than what we found for several (dispersion-corrected) density functionals.
Multi-reference perturbation theories, when applied for single-configurational species, yielded results close to single-reference perturbation theories, and hence, do also not achieve a good agreement with the DLPNO-CCSD(T) data.
We found that M06-L results exhibited the lowest mean absolute deviation (13.2 kJ\,mol$^{-1}$)
out of the results obtained with nine pure density functionals (PBE, BP86, BLYP, TPSS,
M06-L, PBE0, B3LYP, TPSSh, M06-2X).
This may be rationalized by the fact that the Minnesota functionals describe dispersion 
interactions to some degree through their parametrization in contrast to the other 
functionals (PBE, BP86, BLYP, TPSS, PBE0, B3LYP, TPSSh).

Naturally, an adequate description of dispersion interactions is crucial to 
yield accurate \textit{gas-phase} ligand dissociation energies for large transition metal complexes.
The agreement with the DLPNO-CCSD(T) results improved significantly for every density functional
when semiclassical D3 dispersion corrections were considered.
We found that the D3 corrections for the B3LYP functional agreed well
with other density-dependent dispersion corrections (B3LYP-NL, B3LYP-SCNL, B3LYP-XDM) which 
may be taken as further evidence that they are reliable.
We explicitly assessed the uncertainty of the B3LYP-D3(BJ) relative dispersion energies with
\textsc{BootD3}\cite{Weymuth2018} which is only 2.0 kJ~mol$^{-1}$ on average and which 
does not exceed 6.6 kJ~mol$^{-1}$ for the WCCR10 set.

For the whole WCCR10 set, DLPNO-CCSD(T) and all D3-dispersion-corrected density functionals yielded similar results.
The results obtained with PBE0-D3(BJ) agreed very well with the DLPNO-CCSD(T) data.
The PBE0-D3(BJ) results deviated at most by 9.1 kJ~mol$^{-1}$ and on average by 4.3 kJ~mol$^{-1}$
from the DLPNO-CCSD(T) energies.
A system-focused density-functional parametrization in combination with rigorous
error estimation\cite{Simm2016,Simm2017} could reduce these already small errors further
when highly accurate results are required as, for instance, for the elucidation of reaction kinetics. \cite{Proppe2017}

\section*{Acknowledgements}

This work was supported by the Schweizerischer Nationalfonds (Project No. 200020\_169120).
L.F. acknowledges the Austrian Science Fund FWF for a Schr\"{o}dinger fellowship (Project No.\ J 3935).
We are grateful to Prof. Peter Chen for helpful discussions.

\section*{Supporting Information}
Details on parameter-dependence studies, orbital entanglement diagrams, the active space selection for multi-configurational calculations, as well as total electronic eneries and 
DFT and MP2 ligand dissociation energies
can be found in the Supporting Information.
This information is available free of charge via the Internet at http://pubs.acs.org/.

\providecommand{\refin}[1]{\\ \textbf{Referenced in:} #1}


\end{document}